# State Dependent Statistical Timing Model for Voltage Scaled Circuits

Aras Pirbadian$^\xi$, Muhammad S. Khairy$^\dagger$, Ahmed M. Eltawil$^\xi$ and Fadi J. Kurdahi$^\xi$
$^\xi$ Electrical Engineering and Computer Science Department
University of California Irvine, Irvine, CA, USA
$^\dagger$ Qualcomm Corporate R&D, San Diego, CA, USA
Email: $^\xi$ {apirbadi, aeltawil, kurdahi}@uci.edu, $^\dagger$mkhairy@qualcomm.com

*Abstract-* **This paper presents a novel statistical state-dependent timing model for voltage over scaled (VoS) logic circuits that accurately and rapidly finds the timing distribution of output bits. Using this model erroneous VoS circuits can be represented as error-free circuits combined with an error-injector. A case study of a two point DFT unit employing the proposed model is presented and compared to HSPICE circuit simulation. Results show an accurate match, with significant speedup gains.**

## I. Introduction

In advanced CMOS technology, process, voltage and temperature variations (PVT) are becoming significant factors that affect system and circuits reliability. The practice of excessive margining to protect against PVT variations leads to over designed systems that are power inefficient. In fact, the International Technology Roadmaps for Semiconductors (ITRS) has indicated that tradeoffs involving reliability versus energy-efficiency are major challenges for the semiconductor industry [1]. To reduce power consumption, Voltage over Scaling (VoS), defined as extending the voltage scaling range beyond the typical error free regions, has been adopted as an effective means to energy-efficient systems and circuits in advanced CMOS technology.

Several works presented in literature utilize fault tolerant algorithms and circuits to tradeoff system reliability versus energy efficiency [2]-[4]. The main concept behind these systems is to allow errors to occur in a controlled-manner by violating timing constraints due to supply voltage reduction, while maintaining a certain reliability or quality of service dictated by the system. An alternative method to reduce power consumption is by utilizing Dynamic Voltage and Frequency Scaling (DVS) [5]. The main concept behind DVFS is to utilize available time-slack to relax both the supply voltage and frequency of the system while satisfying 100% error-free operations. In both techniques, it is required to accurately model the circuit timing under different supply voltages. In VoS, the percentage of time where the circuits output timing violates the timing constraint defines the error rates in the circuit and hence its reliability. While in DVFS, to guarantee error-free operation, the worst case timing under the reduced supply voltage should be known in order to determine the corresponding reduced clock frequency.

Numerous techniques have been investigated to reflect the effect of VoS at the circuit level to the system QoS. One way to achieve this goal is to adopt a dynamic timing analysis (DTA) [6] approach by integrating SPICE into system level simulation. However, it will be inefficient due to the lengthy processing time. Alternatively, statistical static timing analysis (SSTA) [7] rapidly provides useful statistics of propagation delays and timing violations of critical paths, however it does not provide any information about the overall timing distribution. Though some SSTA work as in [8] considered input vector to exercise false paths, to the best knowledge of the authors, no similar work considered the input vector state dependency for deriving timing distribution. A fast error-aware timing model was presented in [9] based on an enhanced DTA approach. However, since it is a Monte Carlo based approach; it suffers from lengthy processing time for large circuits.

In this paper, we extend the state-dependent model presented in [9] to a statistical model that is capable of rapidly describing the timing distribution of the propagation delays of logic circuits. Furthermore, this model is used to develop an equivalent architecture where erroneous VoS circuits can be replaced with an error-free circuit followed by an error-injector. This equivalent model can be used to reflect the effect of VoS at the circuit level to the system level.

The paper is organized as follows. In Section II, the proposed timing model is described. The error injection model is discussed in Section III. In Section IV, a case study employing the model is analyzed. Finally, the paper is concluded in section V.

## II. Porposed Timing Model

### a) Related Work

It has been shown in [9] that the propagation delay of logic gates under voltage scaling can be accurately approximated by a Gaussian distribution. The mean and variance of this distribution depends on the previous and current states of the inputs to the gate and can be obtained at the characterization phase [9]. Consequently, for a sample gate with n inputs there would be $2^{2n}$ possible input transitions sets which correspond to $2^{2n}$ sets of mean and variance for the Gaussian distribution for every voltage. Furthermore, [9] proposed a method for obtaining the propagation delays of arithmetic circuits such as adders and multipliers under voltage scaling which shows a very close match as compared to HSPICE circuit simulation while achieving a speed up gain by several orders of magnitude. However, this model employs Dynamic Timing Analysis (DTA) combined with a Monte Carlo approach which requires an excessively long time to obtain the timing distribution for large circuits. As an example, for an 8 bit adder, an input vector set of size 4 Giga input samples is required to cover all possible input combinations. This is a major concern for more complex circuits. In this paper, we propose an analytical approach to rapidly obtain accurate timing propagating delays of circuits under voltage scaling and achieve a speed up gain of one to two orders of magnitude as compared to [9].

*b) Proposed Model*

We propose a state dependent analytical statistical model for the propagation delay of logic circuits. The model uses the input transition states coupled with the logic function of the gate to obtain the timing distribution of the output transitions. For an output transition from $i$ to $j$ ($i \rightarrow j$; $i,j \in \{0,1\}$), $T_y^{i \rightarrow j}$ is defined as the timing distribution of output $y$ for the transition from $i$ to $j$. For a simple logic gate with two inputs and one output there are sixteen different input transitions (when considering current state and next state) and four possible transitions at the output. Out of these four output transitions two correspond to $1 \rightarrow 1$ or $0 \rightarrow 0$ changes and therefore result in zero propagation delay. This can be represented by a delta function such as

$$T_y^{0 \rightarrow 0} = \delta(t) , T_y^{1 \rightarrow 1} = \delta(t) \tag{1}$$

The timing distribution of the other two transitions can be modeled as a sum of scaled Gaussian distributions as

$$T_y^{0 \rightarrow 1} = \sum_i \beta_i^{0 \rightarrow 1} \times N(\mu_i, \sigma_i) \tag{2}$$

$$T_y^{1 \rightarrow 0} = \sum_i \beta_i^{1 \rightarrow 0} \times N(\mu_i, \sigma_i) \tag{3}$$

where $N(\mu_i, \sigma_i)$ is the normal distribution with mean $\mu$ and standard deviation $\sigma$ of the $i^{th}$ input transition that will change the output from 0 to 1. Note that the proposed model can be extended to assume non-Gaussian input distributions. $\beta_i^{0 \rightarrow 1}$ represents the probability of occurrence of the $i^{th}$ input change. Using these output timing distributions, the total propagation delay distribution is the weighted sum of all four transitions which can be represented as:

$$T_y = \sum_j \sum_j \alpha_{i \rightarrow j} \times T_y^{i \rightarrow j} \tag{4}$$

where $\alpha_{i \rightarrow j}$ is the weight of the output transition $i \rightarrow j$ which can be obtained via knowledge of the logic function of the gate, probability and number of input transitions that generate the $i \rightarrow j$ transition at the output.

*c) Intermediate transition*

Digital logic circuits are implemented as cascaded chains of individual logic gates. The previous section explained the statistical timing distribution of a standalone logic circuit. In this section, we consider cascaded logic blocks. As an example, consider the two cascaded logic gates blocks shown in Fig. 1. To find the timing distributions of the transitions of $Y_2$ and their corresponding probabilities, the intermediate transition of $Y_2$ (or glitches) need to be considered. Generally, there are two possible scenarios for the output $Y_2$ transition. 1) $Y_2$ does not experience an intermediate state and the change in $Y_2$ is solely dependent on $Y_1$ delay and $F_2$ gate delay, and 2) $Y_2$ experiences an intermediate state due to other input changes prior to settling to a final state due to a change in $Y_1$. In the first case, as explained by the cascaded circuit in Fig. 1 and its timing diagram in Fig. 2, the total delay timing of the output is the addition of the timing delay of both blocks. Equivalently, the timing distribution can be expressed as the convolution of delay timing distributions of both gates. An example showing the second case, where $Y_2$ experiences a glitch, is shown in Fig. 2. The timing delay in this scenario is somehow complicated and will be further explained in the next section.

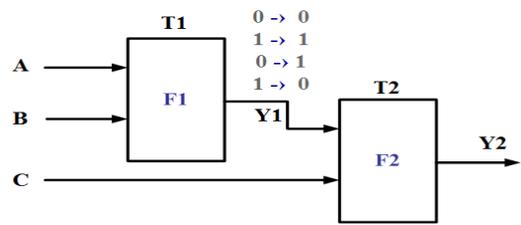

Fig. 1. Cascaded logic gates.

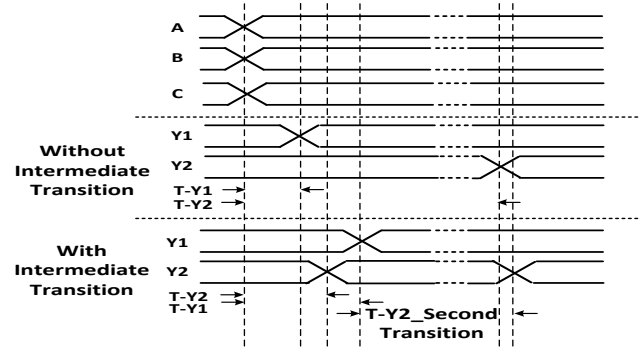

Fig. 2. Timing diagram of logic nets

*d) Simplified example*

To better explain the timing model with intermediate states and without loss of generality, we consider an example of a two XOR circuit where the blocks of Fig. 1 are replaced with XOR gates. Fig. 3 shows a case where the inputs of the logic gate $F_2$ go through a $[(Y_1 C); 11 \rightarrow 00]$ transition. In this figure, $t_1$ is the transition time of C and $t_2$ is the time of the $Y_1$ transition. The transition for $Y_2$ shown in the figure is hypothetical and

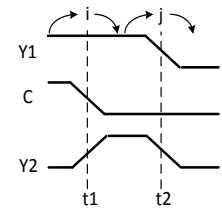

Fig. 3. 11 to 00 transitions for the XOR circuit.

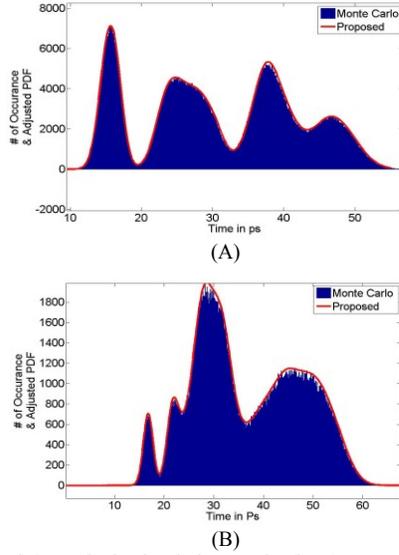

Fig. 4. Hspice/Monte Carlo simulation results (bars) versus proposed statistical model (Solid lines) for (A) XOR circuit on top and (B) adder on bottom

TALBLE I
Run time comparison

|  | XOR | Full Adder | 8-bit ripple adder |
|---|---|---|---|
| Proposed Model | 7 Sec | 10 sec | 25 Sec |
| Monte Carlo[9] | 6 minutes | 8 minutes | 35 minutes |
| Spice | 10 hours | 53 hours | 215 hours |

assumes zero propagation delay in the gate $F_2$. Although the output should not change for such a transition, an intermediate state will occur. To identify the timing distribution for all possible $1 \rightarrow 0$ transitions, including those contributed by glitches, one has to account for the relative delays of the two gates. To do so, the timing distribution of $F_1$ is multiplied with a step function centered at the mean of the timing distribution of $F_2$. Thus the following equation is the contribution of the 11 $\rightarrow$ 00 transition at $F_2$'s input to the output $1 \rightarrow 0$ transition

$$N(\mu_j, \sigma_j) * U(\mu_i) \times T_{y1}^{1 \rightarrow 0} \quad (5)$$

Where $T_{y1}^{1 \rightarrow 0}$ is the timing distribution of the output $Y_1$ transition from $1 \rightarrow 0$ as calculated by (3) and * is the convolution operation. $U(\mu_i)$ is a step function rising at the mean of the second gate's transition and $N(\mu_j, \sigma_j)$ is the $F_2$'s Gaussian timing distribution for the given inputs. In the 11 $\rightarrow$ 00 transition shown in the figure the values $i$ and $j$ represent the indices of the gate $F_2$ input transitions of 11 $\rightarrow$ 10 and 10 $\rightarrow$ 00 respectively.

The total distribution for a given output transition is the addition of the timing distributions of both scenarios and can be expressed as:

$$T_{y2}^{1 \rightarrow 0} = \sum_{i,j} N(\mu_j, \sigma_j) * U(\mu_i) \times T_{y1}^{1 \rightarrow 0} + \sum_i N(\mu_j, \sigma_j) * T_{y1}^{1 \rightarrow 0} \quad (6)$$

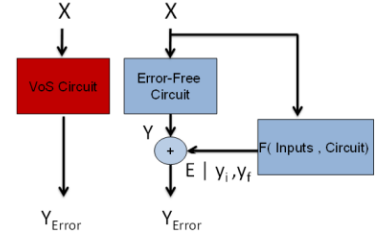

Fig. 5. Equivalent model of erroneous voltage scaled circuit

The pervious equation can be simplified as expressed in (7).

$$T_{y2}^{1 \rightarrow 0} = \sum_{i,j} T_{y1}^{1 \rightarrow 0} * (1 + U(\mu_i)) \times N(\mu_j, \sigma_j) \quad (7)$$

Finally the total output distribution is calculated as in equation (4). Considering that all input values for the XOR circuit are equally likely, then all possible input transitions at $F_2$ will be equally likely and the weights would be equal to 1/16.

*e) Model Verification*

To verify the accuracy of the proposed model, different logic circuits have been tested. We illustrate a comparison between the timing distributions obtained by the Monte Carlo method used in [9] and the proposed statistical model for (a) two cascaded XOR gates and (b) a two-bit adder.

Fig. 4.A illustrates the close match between the timing distribution of propagation delays using the analytical method displayed as a solid line versus the Monte Carlo method, displayed as histogram bars for the XOR circuits. Fig. 4.B provides the results for the last carry bit of a two bit adder. Again, the analytical method accurately predicts the timing distribution at the output. One of the major advantages of the analytical method is the speed up in processing time as shown in Table 1 which illustrates a comparison of the processing time for the HSPICE simulation, Monte Carlo method suggested in [8] and our proposed method for various circuits.

### III. ERROR INJECTION MODEL

The main objective of this work is to find a fast and accurate statistical model for errors when a logic circuit is voltage scaled. Fig. 5 illustrates the equivalent model consisting of an error-free circuit combined with an error injector such that the final output statistics match those of the voltage scaled circuit. The model derives the timing distribution of the output of the voltage scaled logic circuits based on the initial and final values of the input and the output bits. Then, the error is injected by flipping the output bit whenever the signal experiences timing violation.

Consider as an example a 4 bit adder; the delay distribution per each output bit transition is obtained based on the proposed analytical model. The probability of error per output word is then calculated by considering all the possible bit combinations. The errors are then injected at the output of the adder based on the probability of each magnitude of error for the given initial and final values of the output. Fig. 6 shows the probability of each magnitude of error based on our model for a 16 to 0 output transition for 100ps clock frequency.

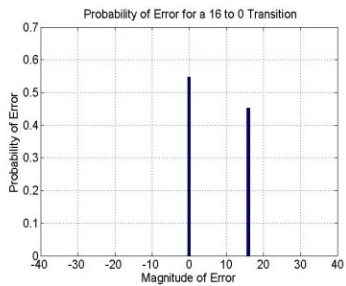

Fig. 6. A sample of the error density function

As shown by the figure the MSB might arrive late due to voltage scaling. The result will be that the output will stay at 16 for the next clock cycle which constitutes an error magnitude of +16 for this transition.

## IV. CASE STUDY

To demonstrate the proposed model at the application level we consider an image compression application (e.g. JPEG) illustrated in Fig. 7 utilizing a 2 point Discrete Fourier Transform (DFT). The 2 point DFT is essentially an addition and subtraction of two inputs as shown in Fig. 8. A sample image is quantized and every 2 pixels are fed into an error free 2-point DFT. The error probabilities are then calculated based on the initial and final values of the DFT output. Error magnitudes are chosen based on the error probabilities calculated by the proposed model. These errors are added to the output of each DFT output. The error free inverse DFT and de-quantization is generated to reconstruct the sample image.

To confirm the accuracy of the proposed method, the same input picture was processed through the Monte Carlo method [9], as well as HSPICE simulations, with 32nm technology and 1V nominal voltage [9]. All tests were performed for a fixed voltage of 70% of the nominal supply voltage (or 0.7V) and a fixed clock period of 100ps. Fig. 9 shows the resulting picture quality for the three methods. To mathematically quantify the quality of the pictures, the peak signal to noise ratio (PSNR) of the output pictures was computed and is displayed below each picture. The processing time for each method is also indicated. The proposed algorithm achieves a speed up gain in processing time by a factor of 60 and 25200 as compared to Monte Carlo [9] and SPICE respectively. On the other hand, the proposed model gives a slightly more pessimistic PSNR result which is appropriate in case the application designer needs to come up with error mitigation algorithms at the application layer (e.g. median filters). In this case there is higher likelihood that these mitigation algorithms will improve the image considerably when implemented.

## V. CONCLUSION

A fast and accurate analytical statistical state-dependent model to calculate the timing distributions of the output bits of voltage scaled logic circuits is presented. The mathematical foundation for the model is explained and its accuracy has been verified. Additionally, an additive statistical error model based on the timing violations has been obtained. Lastly, a case study of a two point DFT employing the model was conducted and the output picture quality in terms of PSNR has been compared to HSPICE circuit simulation.

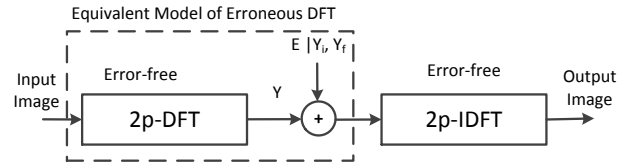

Fig. 7. Block diagram of the case study

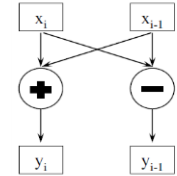

Fig. 8. Two point DFT block diagram

| Proposed | Monte Carlo | HSPICE |
|---|---|---|
| PSNR = 9.94 | PSNR = 10.25 | PSNR = 10.87 |
| Time = 10 sec | Time = 10 mins | Time = 70 Hrs |

Fig. 9. Picture quality, PSNR and timing of HSPICE simulation on right, Monte Carlo in the middle and the proposed method on left.

REFERENCES

[1] International Technology Roadmap for Semiconductors, http://www.itrs.net/.
[2] Lei Wang; Shanbhag, N. R.; "Energy-efficiency bounds for deep submicron VLSI systems in the presence of noise," *IEEE Trans. on VLSI Systems*, vol.11, no.2, pp. 254- 269, Apr. 2003
[3] W. Liu, Renfei; Parhi, K.K.; , "Low-power frequency selective filtering," *Circuits and Systems, 2009. ISCAS 2009. IEEE International Symposium on* , vol., no., pp.245-248, 24-27 May 2009.
[4] Shanbhag, Naresh R.; Abdallah, Rami A.; Kumar, Rakesh; Jones, Douglas L.; , "Stochastic computation," Design Automation Conference (DAC), 2010 47th ACM/IEEE , pp.859-864, 13-18 June 2010
[5] Le Sueur, Etienne, and Gernot Heiser. "Dynamic voltage and frequency scaling: The laws of diminishing returns." *Proceedings of the 2010 international conference on Power aware computing and systems*. USENIX Association, 2010.
[6] Lu Wan and Deming Chen.Analysis of circuit dynamic behavior with timed ternary decision diagram. In Proceedings of the International Conference on Computer-Aided Design (ICCAD '10). IEEE Press, Piscataway,NJ, USA5 16-523.
[7] Blaauw, D.; Chopra, K.; Srivastava, A.; Scheffer, L.; , "Statistical Timing Analysis: From Basic Principles to State of the Art," *Computer-Aided Design of Integrated Circuits and Systems, IEEE Transactions on* , vol.27, no.4, pp.589-607, April 2008.
[8] Jongyoon Jung; Taewhan Kim, "Variation-Aware False Path Analysis Based on Statistical Dynamic Timing Analysis," *Computer-Aided Design of Integrated Circuits and Systems, IEEE Transactions on* , vol.31, no.11, pp.1684,1697, Nov. 2012
[9] Samy Zaynoun, Muhammad S Khairy, Ahmed M Eltawil, Fadi J Kurdahi, Amin Khajeh, "Fast error aware model for arithmetic and logic circuits," *Computer Design (ICCD), 2012 IEEE 30th International Conference on.*, pp.322,328, Sept. 30 2012-Oct. 3 2012.
[10] Predictive Technology Model (PTM). http://www.eas.asu.edu/~ptm